\documentclass[12pt]{article}

\begin{document}
\begin{center}
{\Large \bf  The Role of Conformal Symmetry in Abelian Bosonization
            of the Massive Thirring Model\\}
\vspace*{1cm}
Aleksandar Bogojevi\' c, Branislav Sazdovi\' c\\
{\it Institute of Physics\\
      P.O.Box 57, Belgrade 11001, Yugoslavia\\}
\vspace*{5mm}
Olivera Mi\v skovi\' c\\
{\it Institute of Nuclear Sciences ``Vin\v ca"\\
      Department for Theoretical Physics\\
      P.O.Box 522, Belgrade 11001, Yugoslavia\\}
\end{center}
\begin{abstract}
We show equivalence between the massive Thirring model and
the sine-Gordon theory by gauge fixing a wider gauge
invariant theory in two different ways. The exact derivation of the equivalence hinges on the existence of an underlying conformal symmetry.
Previous derivations were all perturbative in mass (althought to all orders).
\end{abstract}

In a previous paper \cite{bogojevic} we have derived a model which
under two different gauge fixings goes over into the massive Thirring
and sine-Gordon models respectively. Rather than doing this, here we
directly present the wider model. It is given in terms of scalar fields
$\phi$, $\varphi$, spinor field $\psi$ and gauge field $A_\mu$,
living in two dimensional Euclidian space. The generating functional
and Lagrangian are
\begin{eqnarray}
Z&=&\int{\cal D}\bar\psi{\cal D}\psi{\cal D}A{\cal D}\varphi{\cal D}\phi
\,e^{-\int d^2x{\cal L}}\nonumber\\
{\cal L} &=& \bar\psi\gamma_\mu\partial_\mu\psi+
\bar\psi\gamma_\mu\psi \, A_\mu-\frac{1}{2g}\,A_\mu^2+
\frac{1}{2g}(\partial_\mu\varphi)^2+\frac{i}{g}\,A_\mu\partial_\mu\varphi-
\nonumber\\
\label{pocetak}&&-\frac{\pi}{2g}\,(\partial_\mu\phi)^2
-\frac{2\pi}{g\beta}\,\varepsilon_{\mu\nu}\,A_\mu\partial_\nu\phi 
+m\,\bar\psi\psi\cos \beta\phi+im\,\bar\psi\gamma_5\psi\sin \beta\phi\ ,
\end{eqnarray}
where $\frac{4\pi}{\beta^2}=1+\frac{g}{\pi}$. 
We can check that the Lagrangian $\cal L$ and measure
${\cal D}\bar\psi{\cal D}\psi{\cal D}A{\cal D}\varphi{\cal D}\phi$
are invariant under local vector transformations:
\newpage
\begin{eqnarray}
\psi &\to& \psi^\omega=e^{i\omega}\,\psi\nonumber\\ 
\bar\psi &\to& \bar\psi^\omega=e^{-i\omega}\,\bar\psi\nonumber\\
A_\mu &\to& A_\mu^\omega=A_\mu-i\partial_\mu\omega\nonumber\\
\varphi &\to& \varphi^\omega=\varphi-\omega\nonumber\\
\phi &\to& \phi^\omega=\phi\ .
\end{eqnarray}
The invariance of the generating functional follows.
On the other hand, under local axial-vector transformations:
\begin{eqnarray}
\psi &\to& \psi^\lambda=e^{i\lambda\gamma_5}\,\psi\nonumber\\  
\bar\psi &\to& \bar\psi^\lambda=\bar\psi e^{i\lambda\gamma_5}\nonumber\\
A_\mu &\to& A_\mu^\lambda=A_\mu+\varepsilon_{\mu\nu}\partial_\nu\lambda
\nonumber\\
\varphi &\to& \varphi^\lambda=\varphi\nonumber\\
\phi &\to& \phi^\lambda=\phi-\frac{2}{\beta}\, \lambda\ , 
\end{eqnarray}
the Lagrangian and the measure are not invariant. They transform
according to
\begin{eqnarray}
{\cal L}^\lambda &=&{\cal L}+\frac{1}{2\pi}\, (\partial_\mu\lambda)^2
+\frac{1}{\pi}\, \varepsilon_{\mu\nu}A_\mu\partial_\nu\lambda\nonumber\\
({\cal D}\bar\psi{\cal D}\psi)^\lambda&=&{\cal D}\bar\psi{\cal D}\psi\,
\exp \int d^2x\left[ \frac{1}{2\pi}(\partial_\mu\lambda)^2+
\frac{1}{\pi}\varepsilon_{\mu\nu}A_\mu\partial_\nu\lambda\right]\nonumber\\
({\cal D}A{\cal D}\varphi{\cal D}\phi)^\lambda&=&
{\cal D}A{\cal D}\varphi{\cal D}\phi\ .
\end{eqnarray}
The transformation law of ${\cal D}\bar\psi{\cal D}\psi$ is the well
known axial anomaly calculated by Fujikawa \cite{fujikawa}. Taken together,
the effects of the
non invariant terms in the Lagrangian and the measure cancel,
and we are left with an invariant generating functional. $A_\mu$ is an
auxilliary field and integrating it out, we find
\begin{eqnarray}\label{pomocno}
{\cal L} &=& \bar\psi\gamma_\mu\partial_\mu\psi+
\frac{1}{2}\,g\, \left(\bar\psi\gamma_\mu\psi\right)^2+
m\,\bar\psi\psi\cos \beta\phi+im\,\bar\psi\gamma_5\psi\sin \beta\phi+
\nonumber \\
&&+i\bar\psi\gamma_\mu\psi\,\partial_\mu\varphi+\frac{1}{2}\,
(\partial_\mu\phi)^2-\frac{2\pi}{\beta}\,
\varepsilon_{\mu\nu}\,\bar\psi\gamma_\mu\psi\,\partial_\nu\phi\ .
\end{eqnarray}
Now we fix local vector and axial-vector symmetry of $Z$. The first
way to do it is to set $\varphi=0,\enspace \phi=0$. Then  (\ref{pomocno})
becomes
\begin{equation}
{\cal L}_\mathrm{MTM}=\bar\psi(\gamma_\mu\partial_\mu+m)\psi+
\frac{1}{2}\,g\left( \bar\psi\gamma_\mu\psi\right)^2\ .
\end{equation}
This is the famous massive Thirring model \cite{thirring}, a pure fermionic
theory, equivalent to our starting model (\ref{pocetak}).

A second way to gauge fix (\ref{pomocno}) is to take
$\psi_1^\dagger=\psi_1,\enspace
\psi_2^\dagger=\psi_2$, where $\psi=(\psi_1\, , \,\psi_2)^T $.
We then have $\bar\psi\gamma_5
\psi=\bar\psi\gamma_\mu\psi=0$ and (\ref{pomocno}) becomes
\begin{equation}\label{konformna}
\widetilde{\cal L}= \bar\psi\gamma_\mu\partial_\mu\psi+\frac{1}{2}\,
(\partial_\mu\phi)^2+m\,\bar\psi\psi\,\cos\beta\phi\ .
\end{equation}
We haven't obtained a purely bosonic theory despite having
fixed both vector and axial-vector symmetries. However, we still have
at our disposal an additional conformal symmetry. To see this
it is easier look at the Lagrangian (\ref{konformna}) in the operator
formalism, with normal ordered operator fields. Conformal transformations are
given by $z\to z'=f(z)$ and $\bar z\to \bar z'=\bar f(\bar z)$,
where we have introduced complex coordinates $z=x_0+ix_1$ and
$\bar z=x_0-ix_1$. Spinors transform according to
\begin{eqnarray}\label{spinori}
\psi_1(z)\to \psi_1'(z')&=&
\left(\frac{df}{dz}\right)^{-\frac{1}{2}}\psi_1(z)\nonumber\\
\psi_2(\bar z)\to \psi_2'(\bar z')&=&
\left(\frac{d\bar f}{d\bar z}\right)^{-\frac{1}{2}}\psi_2(\bar z)\ .
\end{eqnarray}
The conformal weights can be read off directly from the
correlators for spinor fields $\left\langle\psi_1(z)
\psi_1^\dagger(\zeta)\right\rangle=\frac{1}{2\pi}\,\frac{1}{z-\zeta}$,
and similary for $\psi_2(\bar z)$. From (\ref{spinori}) we easily find
transformation laws of all fermionic terms, writing them in components.
On the other hand, it is known \cite{klaiber} that the free correlator
for a two dimensional massless scalar field is
\begin{equation}\label{korelator}
\left\langle\phi(x)\phi(y)\right\rangle=-\frac{1}{\beta^2}\,
\ln \mu^2(x-y)^2\ ,
\end{equation}
where $\mu$ is an infra red regulator with dimension of mass.
The scale $\mu$ plays a central role. It is, in fact, advantageous
to write the scalar field as $\phi(x|\mu)$.
At the end of all calculations we take the $\mu \to 0$ limit.
From the above correlator we see that $\phi$ transforms in a
complicated way under conformal transformations. However, as is well known,
its derivative has a simple transformation law:
\begin{eqnarray}
\partial_z\phi(x|\mu)&\to&
\partial'_z\phi'(x'|\mu)\,=\,\left(\frac{df}{dz}\right)^{-1}
\partial_z\phi(x|\mu)\nonumber \\
\partial_{\bar z}\phi(x|\mu)&\to&
\partial'_{\bar z}\phi'(x'|\mu)\,=\,\left(\frac{d\bar f}{d\bar z}\right)^{-1}
\partial_{\bar z}\phi(x|\mu)\ .
\end{eqnarray}
Another set of objects made out of $\phi$ transforming in such
a simple way are the normal ordered exponentials of $\phi$.
Using the identity $\left\langle :e^A:\,:e^B:\right\rangle =e^{\left\langle
AB\right\rangle}$ which is valid when $[A,B]$ is a $c$-number, we have
$\left\langle :e^{i\beta\phi(x|\mu)}:\,:e^{i\beta\phi(y|\mu)}:
\right\rangle =0$, and $\mu^2\left\langle :e^{i\beta\phi(x|\mu)}:\,
:e^{-i\beta\phi(y|\mu)}:\right\rangle = \frac{1}{(x-y)^2}$ in
the $\mu\to 0$ limit. As a consequence, for the cosine we get
\begin{equation}\label{kosinus}
\mu\,:\cos\beta\phi(x|\mu):\,\to \,
\left(\frac{df}{dz}\right)^{-\frac{1}{2}}
\left(\frac{d\bar f}{d\bar z}\right)^{-\frac{1}{2}}
\mu\,:\cos\beta\phi(x|\mu):\ .
\end{equation}
Therefore, the whole Lagrangian density transforms like
$\widetilde{\cal L}'(x')=\left(\frac{df}{dz}\right)^{-1}
\left(\frac{d\bar f}{d\bar z}\right)^{-1} \widetilde{\cal L}(x)$ and,
because of $d^2x'\equiv dz'd\bar z'=\frac{df}{dz}\, \frac{d\bar f}{d\bar z}
\, d^2x$, we have the conformal invariant quantum action $\int d^2x
\widetilde{\cal L}$. Fixing the conformal simmetry by $\psi_1=
\theta \left(\frac{df}{dz}\right)^{\frac{1}{2}}, \enspace \psi_2=\bar\theta
\left(\frac{d\bar f}{d\bar z}\right)^{\frac{1}{2}}$, where $\theta$
and $\bar\theta$ are Grassmann constants normalized by $\bar\theta\theta
=-\frac{i\alpha}{2m\beta^2}=$ Const, we find that $m\,\bar\psi\psi =
\frac{\alpha} {\beta^2}=$ Const. Then (\ref{konformna}) becomes
\begin{equation}
{\cal L}_\mathrm{SG}=\frac{1}{2}\, (\partial_\mu\phi)^2+\frac{\alpha}{\beta^2}
\, \cos \beta\phi\ ,
\end{equation}
where the free fermionic Lagrangian was integrated out.
${\cal L}_\mathrm{SG}$ is the well known sine-Gordon model, a pure bosonic
theory.

In this paper we have re-derived Abelian bosonization results
of Coleman \cite{coleman}, Mandelstam \cite{mandelstam} and others
\cite{dorn} - \cite{damgaard}, concerning  the equivalence between the
massive Thirring and sine-Gordon models. Contrary to our derivation, all the previous results were perturbative (to all orders) in mass $m$. As we have seen, the central point in the above equivalence is the existence of {\it two} mass scales $m$ and $\mu$,
and the fact that in (\ref{konformna}) they enter solely through
their ratio $\frac{m}{\mu}$.

\end{document}